\documentclass[10pt,showkeys]{revtex4-2}
\usepackage[driverfallback=dvipdfm]{hyperref}
\usepackage{color}
\usepackage{graphicx}
\usepackage{makeidx}
\usepackage{bm}
\usepackage[title]{appendix} 
\usepackage{tabu}
\usepackage{amssymb}
\usepackage{float}
\usepackage{pbox}

\usepackage[symbol]{footmisc}

\begin{document}
\title{\LARGE Survival prediction of head and neck squamous cell carcinoma using machine learning models}

\author{Saurav Mandal$^1$}
\email[]{Corresponding author: saurav13\_{sit@jnu.ac.in} or saurav.mnnit@gmail.com}
\author{Akshansh Gupta$^2$}
\author{Waribam Pratibha Chanu$^1$}

\affiliation{$^1$School of Computational and Integrative Sciences, Jawaharlal Nehru University, New Delhi-110067, India.}
\affiliation{$^2$CSIR, Central Electronics Research Institute, Pilani, Rajasthan 333031}

\begin{abstract}
{\noindent}\textbf{Abstract} Head and Neck Squamous Cell Carcinoma (HNSCC) is one of cancer type that is most distressing leading to acute pain, effecting speech and primary survival functions such as swallowing and breathing. The morbidity and mortality of HNSCC patients have not significantly improved even tough there has been advancement in surgical and radiotherapy treatments. The high mortality may be attributed to the complexity and significant changes in the clinical outcomes. Therefore, it is important to increase the accuracy of predicting the outcome of cancer survival. Few cancer survival prediction models of HNSCC have been proposed so far. In this study, genomic data (whole exome sequencing) are integrated with clinical data to improve the performance of prediction model. The somatic mutations of every patient is processed using Multifractal Deterended Fluctuation Analysis (MFDFA) algorithm and the parameter values of Fractal Dimension (Dq) is included along with clinical data for cancer survival prediction. Feature ranking proves that the new engineered feature is one of the important feature in prediction model.
In order to improve the performance index of models, hyperparameters were also tuned in all the classifiers considered. 10-Fold cross validation is implemented and XGBoost (98\% AUROC, 94\% precision, and 93\% recall) proves to be best model classifier followed by Random Forest 93\% AUROC, 93\% precision, and 93\% recall), Support Vector Machine (84\% AUCROC, 79\% precision, and 79\% recall)  and Logistic Regression (80\% AUROC, 77\% precision, and 76\% recall). 

\end{abstract}

\maketitle

\vskip 1cm
$  $
\noindent

{\noindent}\textbf{\large 1. Introduction}\\
\vskip 0.3cm
{\noindent}Cancer is becoming the leading cause of death around the world and is increasingly contributing in reducing the life expectancy in every country. The mortality and morbidity are rapidly growing and is becoming a major health crisis. In 2020 an estimate of 19.3 million new cases and 10 million cancer deaths were reported around the world. 58.3 \% of cancer deaths are estimated to occur in Asia, where 59.3\% global population resides\cite{sung}. Europe accounts for 22.8\% followed by 20.9\% in America of the total cancer cases. Compared to other regions the share of cancer deaths in Asia and Africa are higher compared to other regions \cite{sung}.  Cancer mortality rates are increasing worldwide and therefore it is becoming increasingly important for the development of accurate methods for cancer survival predication\cite{1_wang}. 
Head and Neck Squamous Cell Carcinoma (HNSCC) arises in the squamous cells that are flat and thin cells present inside mucosal and moist surfaces of mouth, throat or nose. HNSCC can be further categorized based on the region of origin such as Oral cavity, Hypopharynx, Nasopharynx, Oropharynx, Larynx and Salivary glands\cite{two,three}. HNSCCs are the sixth most common cancers around the world accounting for 355,000 deaths and 633,000 incident cases, characterized by biological, aetiological, phenotypic and clinical heterogeneity\cite{TCGA,three}. Risk factors of HNSCCs include tobacco and alcohol consumption, although the magnitude of the risk factor may vary drastically in each subgroup\cite{TCGA,6_jcm}. HNSCC treatment is based on various combination of radiotherapy, chemotherapy and surgery based on TNM staging and the anatomic site involved\cite{5_jcm}. 
\par Cancer is a heterogeneous disease having different response to therapy, morphology, molecular features and behavior\cite{2_wang}. It becomes extremely challenging to predict and treat cancer patients because of invasiveness and their clinical outcomes showing significant changes\cite{wang}.  Therefore, to increase the life expectancy of cancer patients through accurate decisions and proper therapeutic guidance it becomes important to predict the cancer with accuracy for cancer prognosis\cite{wang}. Cancer prognosis plays an important role in clinical work for proper and accurate survival outcome for clinicians. The clinical decisions are made on the basis of prognostic prediction knowledge by the clinicians to estimate the cancer prognosis more accurately and reasonably\cite{wang,8_wang}. For more practical strategy for prognosis of the complex HNSCC both the clinical feature and the genetic features combined together may provide more information and lead to more accurate prognosis\cite{8_wang}.  
\par  Machine learning encompasses a broad range of algorithms, a sub-field of artificial intelligence, associates the framework of the learning from the given data sample to a general concept of the interference \cite{bish,tm, written}. There is a two-phase of any machine learning algorithm to estimate the unknown dependency in a system from a given data-set and predict the system using estimated dependencies \cite{Kourou}. There are two types of machine learning algorithms, supervised, where training data-set has a class-label, and another one is unsupervised, where no class-label in the training data-set. The performance of any learning algorithm highly depends on feature engineering.
\par Previously, clinical data have been combined with microarray data to accurately predict cancer survival of the patient\cite{8_wang}. In a study Gevaert et al. developed a Bayesian Network Model in breast cancer prediction by integrating information from 70 genes through different strategies (including partial integration, complete and decision-making) and clinical data to prove that by combining the clinical and microarray data the outcome can be drastically improved\cite{wang,14_wang}. Tian et al used mRNA expression-based stemness index (mRNAsi) based signature in predicting HNSCC survival using machine learning algorithms. In HPV positive patients significantly higher number of mRNAsi was observed as compared to HPV negative patients\cite{cai}.  Jong et al. predicted survival probabilities in HNSCC patient based on clinical data such as Tumor, Lymph Node and Metastasis (TNM) stage, prior malignancy, age of diagnosis, gender etc using Cox-regression model\cite{jong}. 
\par In this paper, machine learning models such as Extreme Gradient Boosting (XGBoost), Random Forest, Logistic Regression and Support Vector Machine (SVM) have been implemented by considering the clinical features and the somatic mutations of Whole Exome Sequencing (WXS) from The Cancer Genome Atlas (TCGA) database that will help clinical prognosis with higher accuracy. Long range correlation leads to multifractal properties in the DNA sequences and pseudorandom nucleotide dsitribution\cite{7_man,10_man}. The DNA sequence can be encoded as DNA random walk and through multifractal detrended fluctuations analysis (MF-DFA) technique various parameters such as Hurst exponent(Hq) and Fractal Dimension (Dq) can be calculated. Earlier this powerful technique has been used to distinguish between different types of drug resistant, multi-drug resistant and extremely drug resistant genomic isolates of \textit{Mycobacterium tuberculosis}\cite{ilen,man}.  It has also been applied to study sunspot time	series data\cite{sunspot}, brain EEG data\cite{eeg}, earthquake data\cite{earthquake} etc. To extract useful information from the somatic mutations of HNSCC patients from the TCGA database MFDFA and providing as an input to the machine learning models is a plausible option.

\vskip 0.7cm
{\noindent}\textbf{\large 2. Methods}\\
{\noindent}

\begin{figure}[h!]
\centering
\includegraphics[height=15cm,width=13cm]{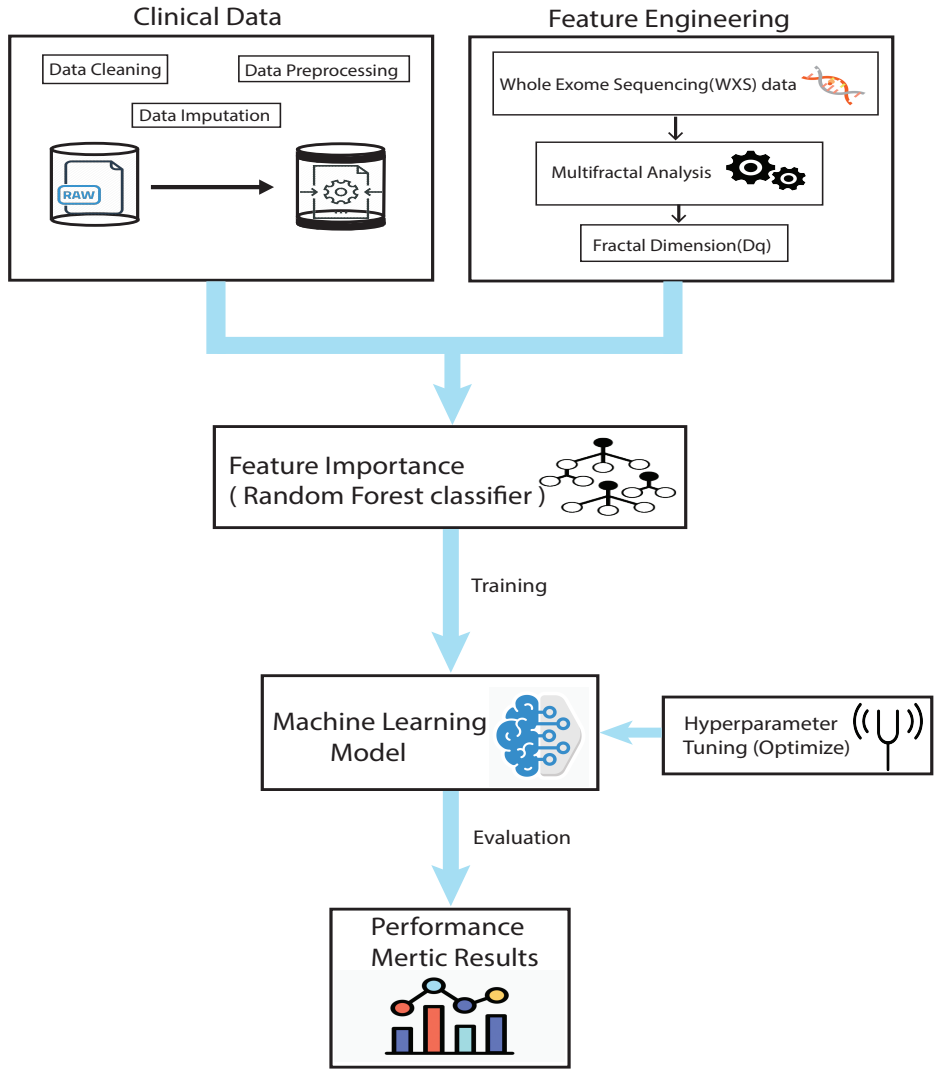}
\caption{ Summary workflow of the machine learning process for cancer survival prediction.}
\label{pipe}
\end{figure}

\vskip 0.3cm
{\noindent}\textbf{ 2.1 Experimental Data} \\
\\
 TCGA is a comprehensive database of various cancer types and maintains the data with high confidence level. It has multiple genomic and clinical data that includes DNA Methylation, WXS data, RNA-seq data, Copy Number Variation (CNV) data etc \cite{TCGA}. 527 HNSCC clinical data available at TCGA database is considered for analysis \cite{TCGA}.  Somatic point mutations identified through WXS extracted from MuTect\cite{mutect} variant caller is collected for analysis as well. 
 \\ 
\vskip 0.1cm
{\noindent}\textbf{ 2.2 Data Pre-processing and Imputation}\\
\\
5 years is taken as thresholds to divide the samples into long-surviving and short-surviving patients. In the classification output label short-survival patients is set as 0 and long-survival patients are set as 1. Out of 527, 336 are short-surviving patients and 191 are long surviving patients. Total number of males are 385 and females are 142. American Joint Committee on Cancer (AJCC) Tumor, Lymph Node and Metastasis (TNM) and other features are converted from categorical to numerical data \cite{ajcc}.  Missing values in AJCC TNM are replaced by referring to AJCC staging manual, the values of TX, NX and MX are replaced with categorical value of 0 that corresponds to inability to access the TNM stage. Total of five records had missing value in AJCC clinical  features. AJCC pathological M feature is dropped because of higher number (271) of missing values. Figure \ref{pipe} explains the experimental workflow.
 \\ \vskip 0.1cm
{\noindent}\textbf{ 2.3 Feature Engineering}\\
\\ Initially, all the somatic mutations present in a particular patient is converted into numeric sequences by considering purine (A and G) as step-up (+1) and pyrimidine (C and T) as step-down (-1)\cite{walk}. MFDFA is then implemented to calculate multifractal parameters such as Hurst exponent (Hq) and Fractal Dimension (Dq).  Dq is calculated from Hurst exponent (Hq), that is a statistical measure of self-similarity. 
 \\ \vskip 0.1cm
{\noindent}\textbf{ 2.3.1 Multifractal Approach} \\
\\ To calculate Multifractal detrended fluctuation analysis we initially converted the concatenated somatic mutations for each patient into DNA random walk by considering purine(A and G) as  where $x_i=+1$ and pyrimidine(T and C) as  where $x_i=-1$. We calculated various multifractal parameters by considering this as a non-stationary time series data such as Hurst exponent($Hq$),generalized dimension($Dq$),fluctuation function($Fq$), singularity spectrum($f$) numerically described by Kantelhardt et al\cite{mf_12, walk}.
Initially the time series signal ${x_i}$ having length of M of the complete concatenated somatic mutations is converted into profile by calculating mean $(\mu)$ of the signal as follows,
\begin{eqnarray}
M(i)=\sum_{j=1}^{i}(x_j- \mu)
\end{eqnarray}
The above profile calculated is divided into non-overlapping segments of equal size$(s)$. We consider $2N_s$ segments by considering counting from both ends of the signal. The number of segments equals,
\begin{eqnarray}
N_s=int{(\frac{N}{s})}
\end{eqnarray}

Then the variance of the profile is calculated as follows,
\begin{eqnarray}
F^2(s,m)=\frac{1}{s}\sum_{i=1}^s\{M[(m-1)s+i]-M_m(i)\}^2
\end{eqnarray}

where, the fitting polynomial segment in a particular segment is given by $m = N_s + 1, ..., 2N_s.$

Then by averaging out over all the segments the $q^{th}$ order fluctuation function is estimated as,
 
\begin{eqnarray}
F_q(s)=\left\{\frac{1}{2N_s}\sum_{m=1}^{2N_s}[M[(m,s)]^{q/2}\right\}^{1/q}
\end{eqnarray}

After calculating the $F_q(s)$ the scaling behaviour of is found out as,
\begin{eqnarray}
F_q(s)\sim s^{H_q}
\end{eqnarray}
where, the measure of correlation properties and self-similarity of the signal is the Hurst exponent $H_q$ and the $H_q$ can be related to the classical scaling exponent $\tau(q)$ as $\tau(q)=qH_q-1$ and by using the singularity exponent  $\alpha=\frac{d\tau}{dq}$. 
The singularity function $f(\alpha)$ can be found out as,
\begin{eqnarray}
f(\alpha)=q\alpha-\tau(q)
\end{eqnarray}	
From equation above the generalized fractal dimension$(D_q)$ can be calculated as,
\begin{eqnarray}
D_q=\frac{\tau(q)}{q-1}
\end{eqnarray}

\vskip 0.1 cm
{\noindent}\textbf{ 2.4 Model Implementation} \\
\\Optimal hyperparameters were searched through GridsearchCV implementation. Tuned parameters were provided as an input to the model classifier implemented through Scikit-Learn\cite{scikit}. k-Fold cross validation method with k=10 was used to increase the reliability and the ability of the model to learn better from the provided datasets\cite{10fold}.

\vskip 1.0 cm
{\noindent}\textbf{\large3. Results and Discussion}\\

\begin{figure}[!ht]
\centering
\includegraphics[height=7cm,width=12cm]{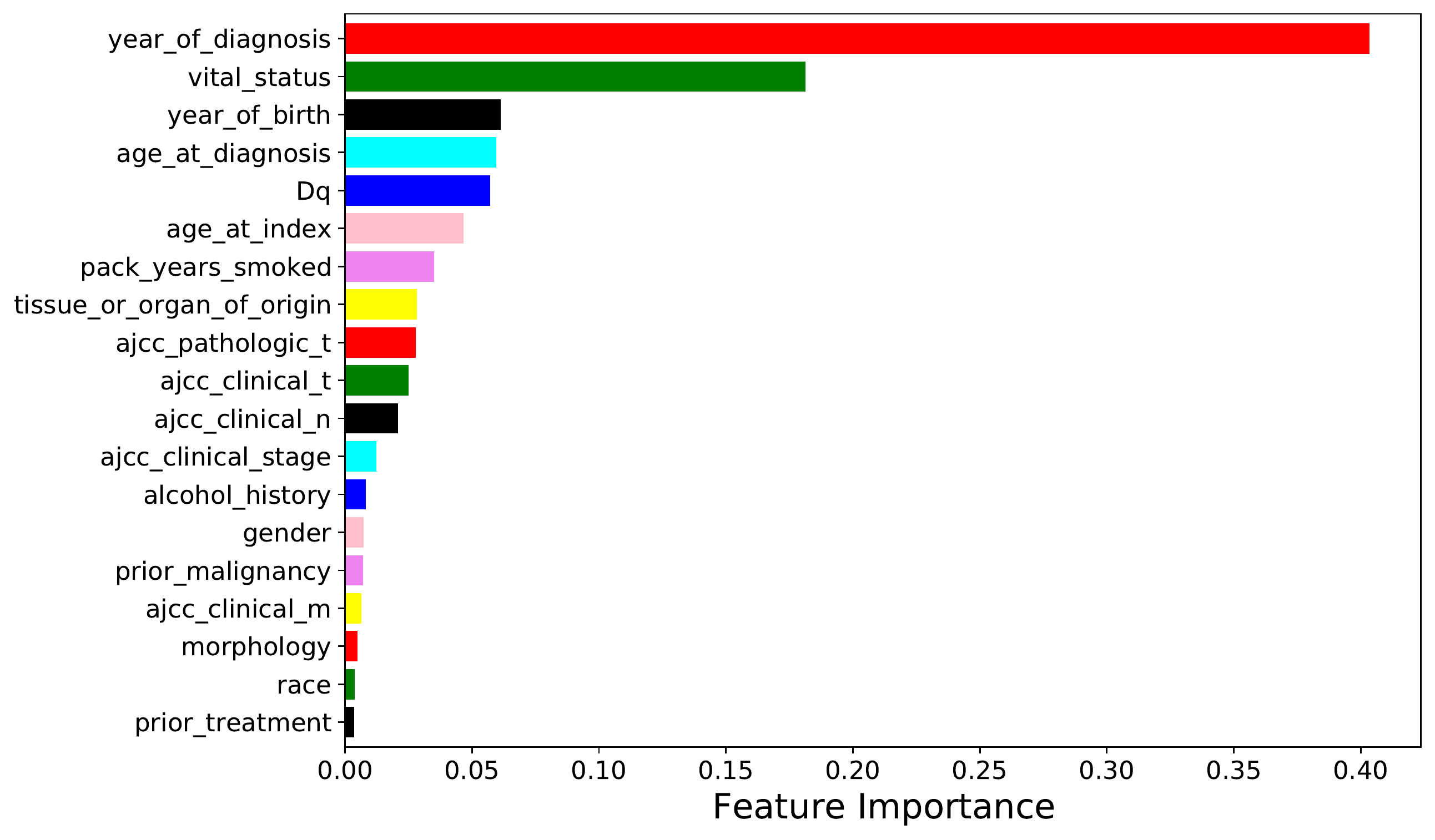}
\caption{Top features based on Random Forest feature importance model.}
\label{features}
\end{figure}

{\noindent}Top features are shown in figure \ref{features}. The features were ranked based on Random Forrest classifier. The most important predictor determined is the year of diagnosis and feature of low importance includes Morphology, race, and prior treatment. Other important predictors are vital status of the cancer patient, year of birth, age of diagnosis, fractal dimension (Dq) etc. Dq is proved to be one of the important features that has been engineered from somatic mutations \cite{TCGA}. Dq is calculated from MFDFA algorithm and has been successfully implemented for classification of drug sensitive, multi drug resistant and extremely drug resistant genomic isolates based on the specific nucleotide sequences\cite{mf_12, man, ilen}. Dq is calculated from Hurst exponent(Hq) that is a statistical measure of self-similarity (Refer material and methods for more details). The workflow of the cancer survival prediction has been represented graphically in figure \ref{pipe}. 

\par Based on the definition of TNM by AJCC the primary tumor(T) can be categorized as TX that is the primary tumor cannot be assessed, T0 being no evidence of primary tumor and Tis means Carcinoma in situ. Similarly, the distribution and the prognostic impact of regional lymph node spread is defined in regional lymph node(N) and presence or absence of distance metastasis information is present in Metastasis(M)\cite{ajcc}.  These features are used in model prediction. The AJCC clinical N feature has  NX, N0, N1, N2, N2a, N2b, N2c and N3 categories where NX refers to the inability to access the lymph node, N0 refers to no regional lymph node metastasis, N1 indicates metastasis in a single ipsilateral lymph node with 3 cm or lesser in greatest dimension, N2 indicates metastasis in a single ipsilateral lymph node or multiple ipsilateral lymph node between 3-6cm, N2a indicates metastasis in a single ipsilateral lymph node with between 3-6cm, N2b refers t metastasis in multiple ipsilateral lymph nodes with none more than 6cm in greatest dimension, N2c refers to metastasis in bilateral or contralateral lymph nodes with none more than 6 cm in greatest dimension and N3 refers to metastasis in a lymph node with more than 6cm in greatest dimension\cite{ajcc}. These eight categorical data of the AJCC clinical N feature were converted into numerical data before training the machine learning model. In HNSCC, the lungs and bones are the most common regions of distant spread of the cancer(metastasis) and brain and hepatic metastases are less frequent. The mediastinal lymph node metastasis are considered as distant metastases. AJCC Clinical Stage feature consists of Stage I, II, III, IVA, IVB, IVC with total number of 21,99, 121, 269, 10 and 7 cases respectively. Total of 27 cases had prior malignancy. 165 cases had no alcohol history and 351 had alcohol exposure. The mean age of diagnosis and pack years smoked are 61.42 and 25.87 respectively. 

\begin{table*}[!h]
\begin{center}

\caption{\label{table:Table1}\bf Hyperparameters}
\begin{tabular}{|p{5cm}|p{6cm}|p{5cm}|p{2cm}|}
 \multicolumn{3}{c}{} \\ \hline

\bf{Model Classifier Name} & \bf{{Hyperparameters and Grid Search Values}}& \bf{Optimized Hyperparameter} 	
\\ 
\hline 
Logistic Regression   &    \pbox{55cm}{C = 0.1, 1, 10\\ Penalty = L1, L2, elasticnet, none\\Solver = newton-cg, lbfgs,\\ liblinear, sag, saga}	&	\pbox{55cm}{C = 0.1,\\ Penalty = L1,\\ Solver = liblinear } 			\\ \hline

Support Vector Machine&  \pbox{55cm}{C = 0.1, 1, 10 \\ kernel = liblinear, poly, rbf\\ gamma = 0.1, 0.01, scale}  &   \pbox{55cm}{ C = 10, \\kernel = rbf,\\ gamma = 0.01}			\\ \hline

Random Forest &  \pbox{55cm}{n\_{estimators} = {100, 1000, 10000} \\ criteria = gini, entropy\\ max\_{depth} = 3, 9, 12}  &   \pbox{55cm}{n\_{estimators} = 100, \\criteria = gini,\\ max\_{depth} = 3}			\\ \hline

XGBoost&  \pbox{55cm}{$\eta$ = 0.01, 0.03, 0.05 \\ max\_{depth} = {3, 6, 9, 12}\\ max\_{child}\_{weight} = {1, 2, 3} }  &   \pbox{55cm}{ $\eta$ = 0.01, \\ max\_{depth} = 3,\\  max\_{child}\_{weight} = 2}			\\ \hline		

\end{tabular}
\end{center}

\end{table*}

\begin{figure}[!h]
\centering
\includegraphics[height=19cm,width=17cm]{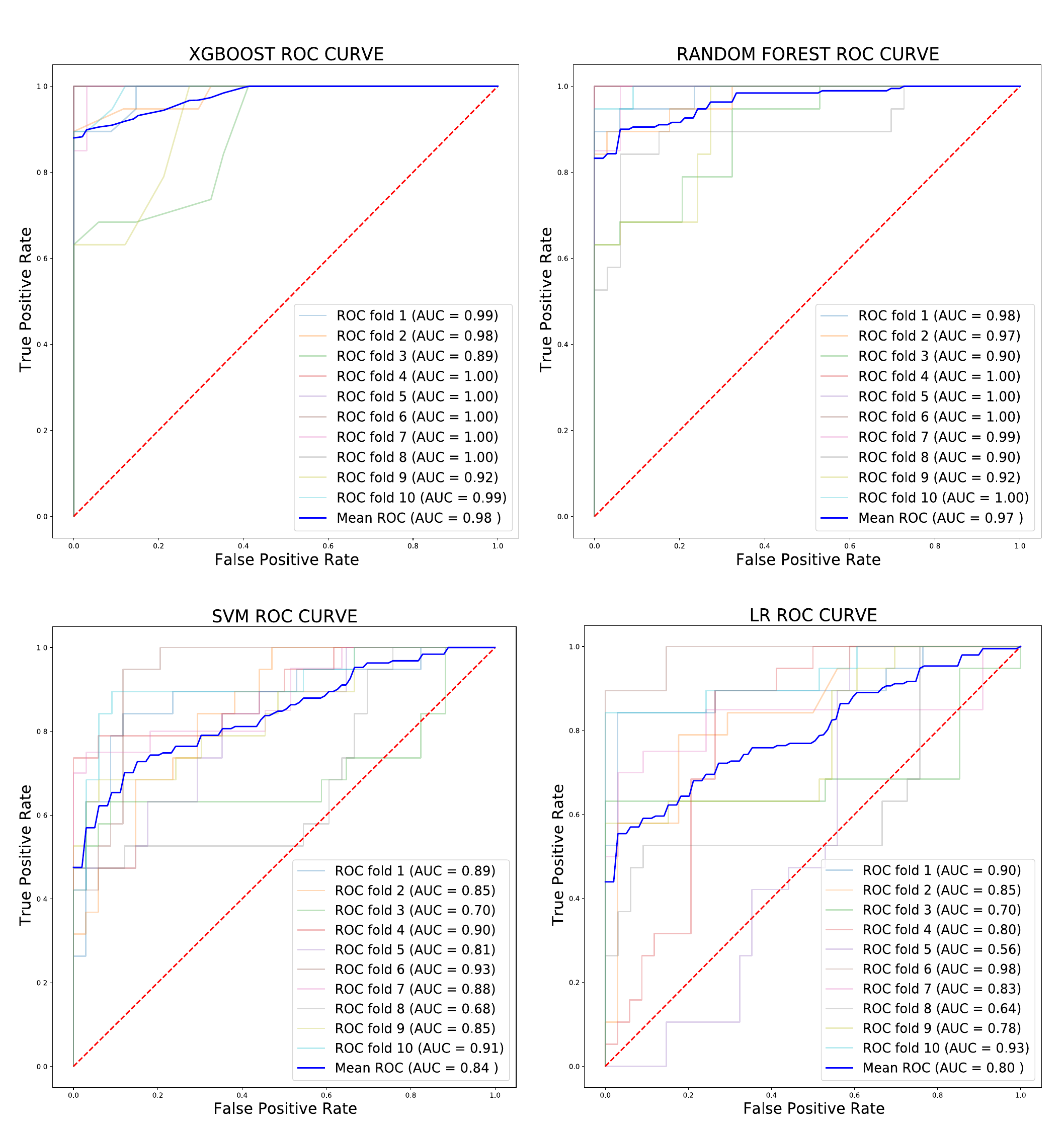}
\caption{ Mean ROC curve shown in blue for XGBoost, Random Forest, SVM and Logistic Regression.}
\label{auc}
\end{figure}
\par The hyperparameters are tuned for each classifier to find optimal hyperparameters as shown in Table \ref{table:Table1}. For XGBoost classifier eta=0.01, max depth = 3,6,9,12 and min child weight = 1,2,3 are given as an input to the perform the grid search and the optimal parameters computed are $\eta$=0.01, max depth = 3 and min child weight = 2(Table \ref{table:Table1}). Similarly, for other classifiers grid search is performed and the optimal hyperparameters are used for model prediction. Performance evaluation indexes-Precision, Recall, F1 Score and AUROC (Mean Values) are evaluated for each machine learning model as described in Table \ref{t2}. XGBoost gives the best performance in terms of precision (0.94), recall (0.93), F1 score (0.93) and AUROC (0.98) followed by Random forest, SVM and Logistic Regression. The metrics are calculated using weighted average and in clinical literature, precision and recall are also called positive predictive value (PPV) and sensitivity (Table \ref{t2})\cite{epi}. It is interesting to observe that the XGBoost and Random forest metrics show that their performance differ by close margin. Also, upon integration of engineered feature fractal dimension (Dq) as a feature the model performance (AUC) increased by 4\% in XGBoost. This indicates that when the somatic mutations Dq value are included along with the clinical data the accuracy of the model tends to perform better for cancer survival prediction. This will help the medical professionals to come to more solid and realistic conclusion regarding the long term or short-term survival chance of HNSCC patient. Thus, based on the performance index of the prediction models proposed upon integration of the somatic mutations information in the form of fractal dimension along with the clinical data will provide a better and concrete prediction results thereby having a positive impact on predicting the outcome.
\\
\par Cross validation is a method that gives model performance with less bias and to avoid problem such as overfitting\cite{10wiki}. In the proposed models k-fold cross validation with k = 10 is implemented. The parameter 'k' in k-fold cross validation indicates that in how many folds the data-set is going to be divided and each will be used in training set (k-1) times. The reliability and the ability of the model to learn from underlying data distribution is better in 10-fold cross validation\cite{2wiki,3wiki}. Receiver operating characteristic (ROC) curve of XGBoost, Random Forest, SVM and Logistic Regression with 10-fold cross validation are shown in Figure \ref{auc}. Mean ROC of XGBoost outperforms other models with AUC of 0.98. Random forest is also observed to perform close to XGboost with AUC 0.97 followed by SVM (AUC=0.84) and LR (AUC=0.80).

\begin{table}[h!]
\begin{center}
\caption{\label{t2}\bf {Model Metrics}}
\begin{tabular}{|l|p{2.5cm}|p{3.5cm}|p{2.5cm}|p{3cm}|}
 \multicolumn{5}{c}{} \\ \hline

\bf{Model Name}& \bf{Precision(PPV)}& \bf{Recall(Sensitivity)} 	& \bf{F1 Score}	& \bf{AUROC(Mean)}	\\ \hline
Logistic Regression & 0.7754 & 0.7666 	&	0.7693 		&0.80	\\ \hline
Support Vector Machine & 0.7935 &  0.7951  	& 0.7951		&0.84	\\ \hline		
Random Forest & 0.9383 &  0.9317  	&0.9299		&0.97	\\ \hline
XGBoost & 0.9430 &  0.9374  &0.9359 & 0.98			\\ \hline
\end{tabular}
\end{center}
\label{table:Table2}
\end{table}
\vskip 1.0 cm
{\noindent}\textbf{\large 4. Conclusion}\\

{\noindent} In this paper the HNSCC patients survival prediction is performed to predict that a patient will be short surviving or long surviving. The somatic mutations were processed using a powerful technique called MFDFA to calculate new engineered feature called fractal dimension. This study improves the prediction accuracy by integrating fractal dimension calculated from somatic mutations of each patient along with their clinical data. The performance index(ROC curve, Recall, Precision, F1 Score) indicate that upon integration the results tend to be more accurate. Hyperparameters were optimized to further improve the performance index of all the model studied. These results may be used in counseling the patient and in making concrete clinical decisions leading to better quality and lifestyle modification required for cancer patient.  
\vspace{1cm}

{\noindent}\textbf{\large Competing interests}\\

{\noindent}Authors declare no competing interests. \\
\vspace{0.3cm}

{\noindent}\textbf{\large Funding statement}\\

{\noindent}SM is financially supported by Indian Council of Medical Research(ICMR) through SRF (Senior Research Fellowship), under  ICMR No. ISRM/11(16)/2019 with submission id 2019/5116. \\

\end{document}